\documentclass{vgtc}                          

\ifpdf
  \pdfoutput=1\relax                   
  \usepackage{graphicx}                
  \DeclareGraphicsExtensions{.pdf,.png,.jpg,.jpeg} 
\else
  \usepackage{graphicx}                
  \DeclareGraphicsExtensions{.eps}     
\fi%

\graphicspath{{figures/}{pictures/}{images/}{./}} 

\usepackage{microtype}                 
\PassOptionsToPackage{warn}{textcomp}  
\usepackage{textcomp}                  
\usepackage{mathptmx}                  
\usepackage{times}                     
\usepackage{cite}                      
\usepackage{tabu}                      
\usepackage{booktabs}                  

\onlineid{0}

\vgtccategory{Research}

\vgtcinsertpkg

\title{Deciphering the Blockchain: A Comprehensive Analysis of Bitcoin's Evolution, Adoption, and Future Implications}

\author{Neelesh Mungoli\thanks{e-mail: nmungoli@uncc.edu}\\ %
        \scriptsize UNC Charlotte %
.} %
       
\abstract{
This research paper provides a comprehensive analysis of Bitcoin, delving into its evolution, adoption, and potential future implications. As the pioneering cryptocurrency, Bitcoin has sparked significant interest and debate in recent years, challenging traditional financial systems and introducing the world to the power of blockchain technology. This paper aims to offer a thorough understanding of Bitcoin's underlying cryptographic principles, network architecture, and consensus mechanisms, primarily focusing on the Proof-of-Work model.

We also explore the economic aspects of Bitcoin, examining price fluctuations, market trends, and factors influencing its value. A detailed investigation of the regulatory landscape, including global regulatory approaches, taxation policies, and legal challenges, offers insights into the hurdles and opportunities faced by the cryptocurrency. Furthermore, we discuss the adoption of Bitcoin in various use cases, its impact on traditional finance, and its role in the growing decentralized finance (DeFi) sector.

Finally, the paper addresses the future of Bitcoin and cryptocurrencies, identifying emerging trends, technological innovations, and environmental concerns. We evaluate the potential impact of central bank digital currencies (CBDCs) on Bitcoin's future, as well as the broader implications of this technology on global finance. By providing a holistic understanding of Bitcoin's past, present, and potential future, this paper aims to serve as a valuable resource for scholars, policymakers, and enthusiasts alike.
} 

\CCScatlist{
  \CCScatTwelve{Bitcoin}{Cry\-pto\-currency}{Block\-chain}{AI}
}

\begin{document}


\firstsection{Introduction}

\maketitle


The introduction of Bitcoin in 2009 marked a turning point in the world of finance, setting the stage for a new era of digital currencies and decentralized technologies. As the first and most prominent cryptocurrency, Bitcoin has experienced significant growth, widespread adoption, and intense scrutiny over the past decade. It has not only captured the attention of the global financial industry but has also spurred the development of numerous other cryptocurrencies and blockchain-based innovations. Despite its continued growth and the increasing prevalence of digital assets, a comprehensive understanding of Bitcoin and its implications remains a topic of great interest and debate.

This paper aims to provide an in-depth analysis of Bitcoin, exploring its historical context, technical foundations, and economic attributes. We begin with a brief overview of the emergence of Bitcoin and the development of blockchain technology, providing essential context for the subsequent discussions. Next, we delve into the cryptographic underpinnings and network architecture that form the backbone of Bitcoin's functionality and security, focusing on the key principles of public and private key cryptography, consensus mechanisms, and the role of miners and mining pools.

Additionally, we examine the economic aspects of Bitcoin, discussing its market dynamics, valuation, and the factors that contribute to its often-volatile price. We also investigate the regulatory landscape surrounding cryptocurrencies, highlighting the varied approaches adopted by different countries, as well as the legal challenges and policy issues that have arisen in relation to Bitcoin.

Furthermore, this paper explores the adoption of Bitcoin across a range of use cases and its impact on the traditional financial system. We discuss the integration of Bitcoin into various financial services, its role in decentralized finance (DeFi), and the opportunities and challenges it presents for businesses and consumers. Finally, we address the future of Bitcoin and cryptocurrencies, identifying emerging trends, technological advancements, and potential challenges, such as environmental concerns and the rise of central bank digital currencies (CBDCs) ~\cite{1}.

By offering a comprehensive analysis of Bitcoin's evolution, adoption, and future implications, this paper seeks to serve as a valuable resource for academics, industry professionals, and enthusiasts alike, fostering a deeper understanding of this groundbreaking technology and its potential to reshape the financial landscape.
\section{Cryptographic Underpinnings and Network Architecture}

\subsection{Public and Private Key Cryptography}
One of the fundamental components of Bitcoin's security and functionality is the utilization of public and private key cryptography. This asymmetric cryptographic system allows users to securely and anonymously transact on the Bitcoin network without the need for a central authority.

A user's private key is a randomly generated number that must be kept secret, as it allows the user to sign transactions and prove ownership of their bitcoins. The public key, on the other hand, is derived from the private key using cryptographic algorithms and can be shared publicly. It serves as an address to which others can send bitcoins. When a transaction is made, the sender uses their private key to create a digital signature, which is then verified by other network participants using the sender's public key. This process ensures that only the rightful owner of the bitcoins can spend them, while the public key provides a transparent and secure means of verifying transactions without revealing the user's identity.
 
\subsection{Consensus Mechanisms: Proof-of-Work and Alternatives}

In order to maintain a decentralized and secure network, Bitcoin employs a consensus mechanism known as Proof-of-Work (PoW). This mechanism requires miners to solve complex mathematical problems to validate and confirm transactions, effectively creating a competitive environment that discourages malicious activities. Once a miner successfully solves the problem, they create a new block of transactions, which is then added to the blockchain. In return, miners receive a block reward in the form of newly minted bitcoins and transaction fees.

Although PoW has successfully secured the Bitcoin network, it has also been criticized for its high energy consumption and potential centralization risks due to the rise of specialized mining hardware. As a result, alternative consensus mechanisms have been proposed and implemented in other cryptocurrencies. Some notable alternatives include:

\begin{itemize}
    \item Proof-of-Stake (PoS): In PoS, validators are chosen to create new blocks and validate transactions based on the amount of cryptocurrency they hold and are willing to "stake" as collateral. This approach reduces energy consumption and encourages long-term investment in the network.
    \item Delegated Proof-of-Stake (DPoS): Similar to PoS, DPoS involves stakeholders voting for a smaller set of trusted validators, creating a more efficient and scalable system.
    \item Proof-of-Authority (PoA): In PoA, a limited number of pre-approved validators, typically chosen for their reputation and trustworthiness, are responsible for validating transactions and maintaining the network.
\end{itemize}

\subsection{The Role of Miners and Mining Pools}
Miners play a crucial role in the Bitcoin network by validating and confirming transactions, as well as securing the blockchain against malicious attacks. They contribute their computational power to solve complex mathematical problems in the PoW process, competing against other miners for the block reward.

Over time, the mining process has become increasingly competitive and resource-intensive, leading to the emergence of mining pools. These pools consist of groups of miners who combine their computational resources to increase their chances of successfully mining a block. When a mining pool successfully mines a block, the reward is distributed among the participants according to their individual contributions, providing a more consistent income for miners.

While mining pools have made it more accessible for individual miners to participate in the mining process, they have also raised concerns about centralization. Some argue that the concentration of mining power within a few large mining pools could potentially compromise the decentralized nature of the Bitcoin network. To mitigate this risk, efforts have been made to develop more decentralized mining pool structures and to encourage the adoption of alternative consensus mechanisms that are less susceptible to centralization ~\cite{2} ~\cite{3} ~\cite{4}.

In conclusion, the cryptographic underpinnings and network architecture of Bitcoin provide the foundation for its secure and decentralized nature. By leveraging public and private key cryptography, consensus mechanisms such as PoW

\section{The Economics of Bitcoin: Market Dynamics and Valuation}

\subsection{Price Fluctuations and Market Trends}
The price of Bitcoin has experienced significant fluctuations since its inception, reflecting the complex interplay of various market forces and investor sentiments. These fluctuations have been driven by a combination of factors, including technological advancements, regulatory developments, macroeconomic trends, and market sentiment. As a result, the price of Bitcoin has seen periods of rapid growth, sharp declines, and periods of relative stability.

Notable market trends include the "halving" events, which occur approximately every four years and reduce the rate at which new bitcoins are created by 50\%. These events have historically been associated with an increase in the price of Bitcoin, as market participants anticipate a reduction in the supply of new coins, leading to higher demand and increased scarcity.

\subsection{Factors Influencing the Value of Bitcoin}
Several factors contribute to the value of Bitcoin and help to explain its price fluctuations. Some key factors include:
\begin{itemize}
    \item Market Sentiment: Investor sentiment plays a significant role in the valuation of Bitcoin. Periods of optimism can lead to rapid price increases, while pessimism can result in sharp declines. Sentiment is often influenced by news events, regulatory developments, and the performance of other cryptocurrencies and financial markets.
    \item Adoption and Use Cases: The adoption of Bitcoin for various use cases, such as payments, remittances, and digital asset management, can influence its value. As more individuals and businesses accept and use Bitcoin, its utility and demand are likely to increase, contributing to a higher valuation.
    \item Regulatory Environment: The regulatory landscape surrounding cryptocurrencies can have a substantial impact on Bitcoin's value. Favorable regulations can encourage adoption and investment, while stricter regulations or enforcement actions can dampen market enthusiasm and hinder growth.
    \item Technological Developments: Advances in blockchain technology, as well as developments in the broader technology sector, can influence the value of Bitcoin. Innovations that improve the scalability, security, or efficiency of the Bitcoin network can increase its appeal, while competition from other cryptocurrencies or technological solutions may pose challenges to its dominance.
    \item Macroeconomic Factors: Broader macroeconomic trends, such as inflation, interest rates, and currency fluctuations, can also impact the value of Bitcoin. Some investors view Bitcoin as a hedge against inflation or economic uncertainty, which can drive demand in times of financial instability ~\cite{5}.
\end{itemize}

\subsection{The Role of Supply and Demand in the Cryptocurrency Market}
The dynamics of supply and demand play a crucial role in determining the value of Bitcoin and other cryptocurrencies. Bitcoin's supply is limited by its algorithmic design, which caps the total number of bitcoins that can ever be created at 21 million. This fixed supply introduces a level of scarcity that can contribute to its value, particularly as the rate of new bitcoin creation decreases over time.

Demand for Bitcoin, on the other hand, is driven by various factors, including its utility as a medium of exchange, a store of value, and an investment asset. As more people and businesses adopt Bitcoin for different use cases, demand is likely to increase, pushing the price higher. Conversely, decreased interest or negative market sentiment can lead to reduced demand and falling prices ~\cite{6}.

The interaction of supply and demand in the cryptocurrency market can result in significant price volatility, as market participants constantly reassess the value of Bitcoin based on new information, changing circumstances, and shifting perceptions. By understanding the complex interplay of these factors, market participants can gain insights into the forces that drive the price of Bitcoin and the broader dynamics of the cryptocurrency market.
\section{Regulatory Landscape and Legal Challenges}

\subsection{An Overview of Global Regulatory Approaches to Bitcoin}
As the popularity and adoption of Bitcoin and other cryptocurrencies have grown, governments and regulatory authorities worldwide have grappled with developing appropriate legal frameworks to govern their use. Regulatory approaches to Bitcoin vary significantly across jurisdictions, reflecting differences in legal traditions, economic priorities, and risk tolerance. Some common regulatory categories include:
\begin{itemize}
    \item Permissive: In permissive jurisdictions, Bitcoin and cryptocurrencies are embraced and encouraged, with minimal regulatory intervention. These jurisdictions often aim to foster innovation and attract investment in the cryptocurrency sector. Examples include Switzerland and Malta.
    \item Balanced: Balanced jurisdictions seek to strike a balance between promoting innovation and mitigating potential risks associated with cryptocurrencies. They typically implement some form of regulation, such as licensing requirements for exchanges or taxation policies, to ensure consumer protection and financial stability. Examples include the United States, the European Union, and Japan.
    \item Restrictive: Restrictive jurisdictions impose strict regulations on cryptocurrency activities or ban them altogether, often citing concerns related to financial stability, consumer protection, or illicit activities. Examples include China and India, although India's stance has been subject to ongoing debate and legal challenges.
\end{itemize}

\subsection{Taxation and Anti-Money Laundering (AML) Policies}
One of the primary areas of regulatory focus for cryptocurrencies is taxation and anti-money laundering (AML) policies. Most jurisdictions require individuals and businesses to report their cryptocurrency holdings and transactions for tax purposes, although specific tax treatments may vary based on factors such as the nature of the transaction, the jurisdiction, and the taxpayer's individual circumstances.

In addition to taxation, regulators worldwide have implemented AML and combating the financing of terrorism (CFT) policies for cryptocurrency-related activities. These policies often require cryptocurrency exchanges, wallet providers, and other service providers to implement Know Your Customer (KYC) procedures, conduct ongoing customer due diligence, and report suspicious activities to relevant authorities. The goal of these policies is to prevent the use of cryptocurrencies for illicit purposes, such as money laundering, terrorist financing, or tax evasion.

\subsection{Notable Legal Cases Involving Bitcoin and Cryptocurrencies}
Several high-profile legal cases involving Bitcoin and cryptocurrencies have shaped the regulatory landscape and raised critical questions about their legal status, classification, and treatment. Some notable cases include:

\begin{itemize}
    \item Silk Road: The Silk Road was an online marketplace that operated on the dark web and facilitated illegal transactions using Bitcoin as its primary currency. The founder, Ross Ulbricht, was arrested in 2013 and later convicted on multiple charges, including money laundering, computer hacking, and conspiracy to traffic narcotics. The case brought the potential use of Bitcoin for illicit activities to the forefront of regulatory concerns.
    \item Mt. Gox: Once the world's largest Bitcoin exchange, Mt. Gox collapsed in 2014 after a massive security breach resulted in the loss of around 850,000 bitcoins. The case led to increased regulatory scrutiny of cryptocurrency exchanges and highlighted the need for consumer protection measures and security standards in the industry.
    \item SEC v. Kik Interactive: In this case, the U.S. Securities and Exchange Commission (SEC) filed a lawsuit against Kik Interactive, alleging that the company's 2017 initial coin offering (ICO) constituted an unregistered securities offering. The case underscored the regulatory challenges surrounding ICOs and the need for clear guidance on the classification of digital assets under securities laws.
\end{itemize}

In conclusion, the regulatory landscape for Bitcoin and cryptocurrencies remains a complex and evolving area, with varying approaches adopted by different jurisdictions. As the industry continues to mature, ongoing legal challenges and regulatory developments will shape the future of cryptocurrencies, influencing their adoption, use cases, and potential impact on the global financial system ~\cite{7} ~\cite{8}.
\section{Bitcoin Adoption: Use Cases and Impact on Traditional Finance}

\subsection{Consumer and Merchant Adoption}
Bitcoin's increasing adoption as a payment method by both consumers and merchants has contributed significantly to its growth and mainstream recognition. Many businesses, ranging from small local retailers to global corporations, have begun accepting Bitcoin as a form of payment. This acceptance is driven by various factors, such as reduced transaction fees, faster settlement times, and the ability to reach a global customer base without the limitations of traditional banking systems ~\cite{8} ~\cite{9}.

Consumer adoption has also been on the rise, as more individuals recognize the potential benefits of using Bitcoin for transactions, remittances, and even as a store of value. Additionally, the proliferation of user-friendly wallet applications and payment services has facilitated greater accessibility and ease of use, further driving consumer adoption.

\subsection{Integration of Bitcoin in Financial Services}
The integration of Bitcoin into various financial services has further expanded its use cases and increased its appeal to a wider range of market participants. Some notable examples of Bitcoin's integration into traditional finance include:
\begin{itemize}
    \item Investment Products: A growing number of financial institutions offer Bitcoin investment products, such as exchange-traded funds (ETFs), futures contracts, and options, allowing investors to gain exposure to Bitcoin without directly owning the underlying asset. These products provide additional avenues for institutional and retail investors to participate in the Bitcoin market and contribute to its overall liquidity and price discovery.
    \item Lending and Borrowing: Several platforms now offer lending and borrowing services using Bitcoin as collateral, providing users with access to credit without the need for traditional banking intermediaries. These services enable individuals and businesses to leverage their Bitcoin holdings for various financial purposes, such as liquidity management or investment opportunities.
    \item Payment and Remittance Services: Payment processors and remittance companies have integrated Bitcoin into their platforms, offering users the ability to send and receive payments in Bitcoin or convert between Bitcoin and fiat currencies. This integration has the potential to lower transaction costs, increase efficiency, and enable access to financial services for unbanked or underbanked populations.
\end{itemize}

\subsection{Decentralized Finance (DeFi) and the Role of Bitcoin}
Decentralized finance (DeFi) has emerged as a rapidly growing sector within the cryptocurrency ecosystem, leveraging blockchain technology to create a wide range of financial services and applications without the need for centralized intermediaries. While most DeFi applications have been built on the Ethereum blockchain, Bitcoin's role in the DeFi space has been growing, with several projects aiming to bridge the gap between the two ecosystems.

Some key developments in the intersection of Bitcoin and DeFi include:
\begin{itemize}
    \item Wrapped Bitcoin (WBTC): Wrapped Bitcoin is an ERC-20 token pegged to the value of Bitcoin, allowing users to utilize their Bitcoin holdings within the Ethereum-based DeFi ecosystem. This tokenization enables Bitcoin holders to participate in various DeFi activities, such as lending, borrowing, and yield farming, without converting their Bitcoin to Ether or other Ethereum-based tokens.
    \item Cross-Chain Bridges: Several projects are working on creating cross-chain bridges between the Bitcoin and Ethereum networks, allowing for seamless movement of assets between the two ecosystems. These bridges aim to facilitate greater interoperability between different blockchains, enabling users to access a wider range of DeFi applications and services using their Bitcoin holdings.
    \item Native Bitcoin DeFi Platforms: A growing number of DeFi platforms are being built natively on the Bitcoin blockchain, leveraging its security and network effects to create decentralized financial services specifically designed for Bitcoin users. These platforms aim to expand the DeFi ecosystem beyond Ethereum and offer Bitcoin holders direct access to various financial services without the need for intermediaries or tokenization.
\end{itemize}

In conclusion, the growing adoption of Bitcoin in various use cases and its increasing integration into traditional financial services and the DeFi ecosystem has had a significant impact on the financial landscape~\cite{11} ~\cite{12}.

\section{The Future of Bitcoin and Cryptocurrencies}

\subsection{Emerging Trends and Technological Innovations}
As the cryptocurrency ecosystem continues to evolve, several emerging trends and technological innovations are shaping the future of Bitcoin and other digital assets. Some key developments include:
\begin{itemize}
    \item Layer 2 Solutions: To address the scalability limitations of the Bitcoin network, several Layer 2 solutions, such as the Lightning Network, are being developed. These solutions aim to increase transaction throughput and reduce fees by creating off-chain payment channels that only interact with the main blockchain when necessary, enabling faster and more efficient transactions.
    \item Privacy Enhancements: Enhancing privacy features within the Bitcoin network has become a priority for many developers, with projects such as Taproot and Schnorr signatures focusing on improving the privacy and security of transactions. These innovations aim to make it more difficult for third parties to link transactions to specific users, increasing the overall privacy and fungibility of Bitcoin.
    \item Smart Contracts and Programmability: Although the Bitcoin network has limited programmability compared to platforms like Ethereum, there are ongoing efforts to expand its capabilities in this area. Projects such as RSK (Rootstock) and Miniscript aim to bring smart contract functionality to the Bitcoin ecosystem, potentially enabling a wide range of decentralized applications and financial services.
\end{itemize}

\subsection{Environmental Concerns and the Shift Towards Sustainable Blockchain Solutions}
Environmental concerns surrounding the energy consumption of the Bitcoin network and its reliance on energy-intensive Proof-of-Work (PoW) consensus mechanisms have been a major point of debate. These concerns have prompted increased interest in more sustainable blockchain solutions, such as Proof-of-Stake (PoS) or other energy-efficient consensus mechanisms.

While it remains uncertain whether the Bitcoin network will adopt a different consensus mechanism in the future, there is growing awareness of the need for more sustainable practices in the mining industry. This includes a shift towards renewable energy sources, improvements in mining hardware efficiency, and the development of carbon offset initiatives.

\subsection{The Potential Impact of Central Bank Digital Currencies (CBDCs) on Bitcoin's Future}
The growing interest in Central Bank Digital Currencies (CBDCs) presents both opportunities and challenges for the future of Bitcoin and cryptocurrencies. CBDCs are digital representations of a nation's sovereign currency, issued and managed by central banks. Their introduction could have several implications for the cryptocurrency ecosystem:

\begin{itemize}
    \item Legitimization: The development of CBDCs could legitimize the concept of digital currencies in the eyes of the public and encourage further adoption of cryptocurrencies, including Bitcoin.
    \item Competition: CBDCs may compete with Bitcoin and other cryptocurrencies for market share, particularly in areas such as cross-border transactions, remittances, and digital payments. However, the decentralized nature of Bitcoin and its inherent properties as a store of value could help it maintain its appeal despite the emergence of CBDCs.
    \item Regulatory Pressure: The introduction of CBDCs could lead to increased regulatory pressure on cryptocurrencies, as governments and central banks seek to maintain control over the monetary system and prevent potential threats to financial stability. This could result in stricter regulations or enforcement actions targeting the cryptocurrency ecosystem.
\end{itemize}

In conclusion, the future of Bitcoin and cryptocurrencies is influenced by a multitude of factors, including emerging trends, technological innovations, environmental concerns, and the potential impact of CBDCs ~\cite{16}. As these forces continue to shape the cryptocurrency landscape, the ongoing development and adoption of Bitcoin and other digital assets will likely transform the way we think about money, finance, and the global economy.

\section{Conclusion}

As we have explored throughout this paper, Bitcoin and cryptocurrencies have come a long way since the inception of the Bitcoin network in 2009. From humble beginnings as an experimental digital currency, Bitcoin has grown into a global phenomenon that has disrupted traditional financial systems and introduced novel concepts such as decentralization, digital scarcity, and programmable money.

The rise of Bitcoin and cryptocurrencies can be attributed to their unique properties and advantages, such as reduced transaction costs, increased financial inclusivity, and the ability to operate beyond the control of centralized institutions. These characteristics have fueled a diverse range of use cases, from digital payments and remittances to investment products and decentralized finance applications ~\cite{12}.

However, the growth of the cryptocurrency ecosystem has also been accompanied by various challenges and concerns, including regulatory uncertainty, environmental sustainability, and the potential for illicit activities. As governments and regulatory bodies worldwide grapple with these issues, the development of appropriate legal frameworks and industry best practices will be crucial in ensuring the long-term viability and success of cryptocurrencies ~\cite{13}.

Innovation and advancements in blockchain technology have driven significant improvements in the scalability, privacy, and programmability of the Bitcoin network and other cryptocurrencies. These ongoing innovations, along with the growing interest in sustainable blockchain solutions and the potential impact of central bank digital currencies, will undoubtedly continue to shape the future of the cryptocurrency ecosystem ~\cite{14} ~\cite{5} ~\cite{6}.

In conclusion, the story of Bitcoin and cryptocurrencies is one of constant evolution, driven by a diverse and passionate community of developers, entrepreneurs, investors, and users. As the cryptocurrency landscape continues to mature and overcome the challenges it faces, the potential for Bitcoin and other digital assets to transform the global financial system and revolutionize the way we conduct transactions, store value, and access financial services seems increasingly likely ~\cite{16}. The future of money is being reimagined, and cryptocurrencies like Bitcoin are at the forefront of this exciting transformation.


\bibliographystyle{abbrv-doi}

\bibliography{template}
\end{document}